\documentclass[10pt, aps, prl,  noshowkeys, 
               twocolumn, floatfix, preprintnumbers]{revtex4-1}
\usepackage{amsmath, amssymb, amsfonts}
\usepackage{graphicx}
\usepackage{color}
\usepackage{times}
\usepackage{multirow}
\newcommand{\bk}{{\bf{k}}}

\newcommand{\bp}{{\bf{p}}}
\newcommand{\bq}{{\bf{q}}}

\newcommand{\bA}{{\bf{A}}}

\renewcommand{\P}{\hat{\Psi}}

\newcommand{\br}{{\bf{r}}}

\begin{document}
\title{Optically bright $p$-excitons indicating strong Coulomb coupling in transition-metal dichalcogenides}

\author{T. Stroucken}
\affiliation{Department of Physics and Material Sciences Center, 
Philipps University Marburg, Renthof 5, D-35032 Marburg, Germany}

\author{S.W. Koch}
\affiliation{Department of Physics and Material Sciences Center, 
Philipps University Marburg, Renthof 5, D-35032 Marburg, Germany}

\begin{abstract}
It is shown that the strong Coulomb coupling in intrinsic suspended semiconducting transition metal dichalcogenides
can exceed the critical value needed for an excitonic ground state. The dipole-allowed optical excitations then 
correspond to intra-excitonic transitions such that the optically bright excitonic transitions near the Dirac points  
have a $p$-like symmetry whereas the $s$-like states are dipole forbidden. The large intrinsic coupling strength
seems to be a generic property of the semiconducting transition metal dichalcogenides and strong Coulomb-coupling signatures in the form of the optical selection rules can be observed
even in samples grown on typical substrates like SiO$_2$. 
For the examples of WS$_2$ and WSe$_2$, excellent agreement of the 
computed excitonic resonance energies with recent experiments is demonstrated.
\end{abstract}
\date{\today}
\pacs{
71.35.Cc 	
78.20.-e,	
}
\maketitle

\section{Introduction}

Monolayers of transition metal dichalcogenides (TMDs) are novel two-dimensional (2D) semiconductors.
Similar to graphene, these
materials display exciting new physical properties, distinct from their
bulk counterparts\cite{Mak2010,Terrones2013,Splendiani2010,Cappelluti2013,Lambrecht2012,Liu2013,Xiao2012,Molina-Sanchez2013,
Cao2012,Zeng2012,Mak2012,Zeng2013,
Chernikov2014,Ye2014}. The linear optical spectra are dominated by strong resonances with  
peak absorption values of more than 10\% of the incoming light. Furthermore, the absence of 
inversion symmetry 
allows for efficient second harmonic generation (SHG). 
Though initially assigned to free-particle transitions, meanwhile the 
excitonic nature of the strong optical response is widely accepted. 
On the basis of recent linear and nonlinear optical experiments 
on WS$_2$ [1-3], the respective authors report large binding energies for 
the energetically lowest, optically bright excitons 
ranging from 0.32-0.71 eV. At the same time, strong deviations from the usual hydrogenic Rydberg series are observed. Due to the difficulty to directly measure the single-particle bandgap energy, 
the binding energies have been deduced either from the energetic separation between different resonances 
or by the comparison with theoretical predictions. 

In general, the proper analysis of the measured spectra  
depends critically on the correct identification of the optically active states. For this purpose,
one derives optical selection rules that result from the conservation of the total angular momentum 
in the excitation and emission processes. The $\Gamma$-point transitions in
conventional direct-gap GaAs-type semiconductors occur between the $sp^3$-hybridized near-parabolic $s$-like 
conduction and $p$-like valence bands. The
near-bandgap optical properties are well described by the so-called Elliot formula\cite{HaugKoch}, 
expressing the energetic position and oscillator strength of the excitonic transitions 
in terms of the excitonic wave functions. As a result of the simple dipole selection rules, 
only $s$-like excitonic states couple to the light field. This 
leads to the well-known excitonic Rydberg series where the lowest optical active state is the 1s 
exciton resonance.

The derivation of the Elliot formula\cite{HaugKoch} is based on 
the implicit assumption that the 
system is excited from the noninteracting ground state.
In contrast, if one considers an excitonic 
insulator\cite{Keldysh1965,Keldysh1968,Jerome1967,Halperin1968,Littlewood1996}, the ground state itself is excitonic and light absorption or emission involves
intra-excitonic transitions
that are governed by fundamentally different optical selection rules. The energetically
lowest transition from an $1s$-type excitonic ground state will lead to optically excited $p$-type excitons.  

In this paper, we present strong evidence that not only for suspended TMDs, but also for TMDs on a SiO$_2$ substrate, the optically active states 
correspond to the $p$-type excited states of a two-dimensional (2D) hydrogenic Rydberg series, 
while the lowest lying $1s$-exciton is merged with the ground-state level.
Our predictions are based on the analysis of the Hamiltonian for 2D massive Dirac Fermions with Coulomb interaction. 
To support our theory, we compare its predictions with recent experiments on 
WS$_2$ \cite{Chernikov2014,Zhu2014,Ye2014} and WSe$_2$\cite{Wang2014}, 
demonstrating that the assumption of bright $p$-states allows for a simple 
interpretation of a wide range of different experimental results. 

\section{Microscopic Theory}

Starting point of our analysis is the low-energy Hamiltonian for the bandstructure in the vincinity 
of the Dirac-points.
Dictated by the symmetry properties of the hexagonal lattice, 
the lowest order $\bk\cdot\bp$ Hamiltonian has the form\cite{Xiao2012}
\begin{equation}
H_0 = \sum_{\tau,\bk}\P^\dagger_{\tau\bk} \left( at{\bk} \cdot {\hat{\bf
\sigma}}_\tau+\frac{\Delta}{2}\hat\sigma_z-\tau\lambda\frac{\hat\sigma_z-1}{2}
\hat s_z \right)\P_{\tau\bk},
\label{Hamiltonian}
\end{equation}
where $\tau=\pm 1$ is the so-called valley index identifying the two non-equivalent Dirac points,
and $\P^\dagger_{\tau\bk}$ is the tensor product of the  electron spin state and
a  two-component quasi-spinor. The Pauli matrices  ${\hat{\bf
\sigma}}_\tau=(\tau\hat\sigma_x,\hat\sigma_y)$ and $\hat\sigma_z$ act in the pseudo-spin space and $\hat s_z$ in the real 
spin space, respectively.
The basis functions for the pseudo-spinors are a linear combination of the relevant atomic orbitals that contribute to 
the valence and conduction bands and depend on the specific material system under consideration.
Within a tight-binding model, the parameters $\Delta$, $t$ and $a$ correspond to the bias in on-site energies, 
the effective hopping matrix element, and the lattice constant, respectively. Furthermore, $2\lambda$ is the spin-splitting 
of the valence band due to the intra-atomic spin-orbit coupling (SOC). 
The Hamiltonian (\ref{Hamiltonian}) is valid for the entire class of monolayer hexagonal structures including
graphene with a zero gap and negligible SOC, h-BN with a large gap and also negligible SOC, and the variety of 
TMDs with a gap in the optical range and strong SOC. 

Independent of the explicit expressions for the pseudo-spinor basis functions, 
the Hamiltonian describing the light-matter (LM) interaction can be found by the minimal substitution $\hbar \bk\rightarrow\hbar\bk-e\bA/c$. Defining the Fermi velocity by $ v_F=at/\hbar$, one obtains
\begin{equation}
\label{HIspinor}
H_I=-e\frac{v_F}{c}\sum_{\tau,\bk}\P^\dagger_{\tau\bk} {\bA} \cdot {\hat{\bf
\sigma}}_\tau\P_{\tau\bk}.
\end{equation}

Using the eigenstates of $H_0$, the LM Hamiltonian can be written as
\begin{widetext}
\begin{eqnarray}
H_I&=&-e\frac{v_F}{c}\sum_{s\tau\bk}\frac{\hbar v_F k}{2\epsilon_{s\tau\bk}}
\left({\rm e}^{-i\tau\theta_{\bk}} A^\tau+{\rm e}^{i\tau\theta_{\bk}}A^{-\tau}\right)
\left(c^\dagger_{s\tau\bk}c_{s\tau\bk}-\nu^\dagger_{s\tau\bk}\nu_{s,\tau,\bk}\right)\nonumber\\
&+&e\frac{v_F}{c}\sum_{s,\tau,\bk}\frac{1}{2\epsilon_{s\tau\bk}}\left\{
\left((\epsilon_{s\tau\bk}-\frac{\Delta_{s\tau}}{2}){\rm e}^{-i\tau\theta_{\bk}}A^\tau
-(\epsilon_{s\tau\bk}+\frac{\Delta_{s\tau}}{2}){\rm e}^{i\tau\theta_{\bk}}A^{-\tau}\right)
 c^\dagger_{s\tau\bk}\nu_{s\tau\bk}
+h.c.\right\} .
\label{HIbands}
\end{eqnarray}
\end{widetext}
Here, $c^\dagger_{s\tau\bk}$ ($\nu^\dagger_{s\tau\bk}$) creates an electron with spin $s$ and valley index 
$\tau$ in the conduction (valence) band, $\Delta_{s\tau}=\Delta-s\tau\lambda$ is the spin and valley
dependent gap, $\epsilon_{s\tau\bk}=\sqrt{\left(\frac{\Delta_{s\tau}}{2}\right)^2+(\hbar v_Fk)^2}$ is the 
relativistic dispersion of a quasi-particle with  rest 
energy $m_{s\tau}v_F^2=\Delta_{s\tau}/2$, $A^\tau=A_x+i\tau A_y$, and $\theta_\bk$ is the angle in $k$-space
defined by ${\rm tan}\theta_{\bk}=\frac{k_y}{k_x}$.

In Eq. (\ref{HIbands}), the first line describes the intraband and the second line the interband transitions, respectively. 
Contributions proportional to $ A^\tau$  correspond to the absorption of a photon with circular polarization 
$\sigma^\tau$.  In zero-gap materials like graphene or silicene, the transition matrix elements for both intra- and 
interband absorption are equal in magnitude. Differences in the transition probabilities exclusively result 
from the different occupation numbers in the initial state. In contrast, 
in a wide gap system, the 
leading orders of the transition matrix elements are given by  $(\epsilon_{s\tau\bk}-\frac{\Delta_{s\tau}}{2})/2\epsilon_{s\tau\bk}\approx (\hbar v_F k)^2/\Delta_{s\tau}^2$,
$\hbar v_F k/2\epsilon_{s\tau\bk}\approx \hbar v_Fk/\Delta_{s\tau}$, and  $(\epsilon_{s\tau\bk}+\frac{\Delta_{s\tau}}{2})/2\epsilon_{s\tau\bk}\approx 1$.
Keeping only contributions up to first order in $\hbar v_F k/\Delta$
gives

\begin{eqnarray}
H_I&\approx&
-\frac{1}{c}\sum_{s\tau\bk}{\bf A}\cdot{\bf j}_{s\tau\bk}\left(c^\dagger_{s\tau\bk}c_{s\tau\bk}-\nu^\dagger_{s\tau\bk}\nu_{s\tau\bk}
\right)
\nonumber\\
&-&
e\frac{v_F}{c}\sum_{s\tau\bk}\left\{ {\rm e}^{i\tau\theta_{\bk}} A^{-\tau}
 c^\dagger_{s\tau\bk}\nu_{s\tau\bk}
+h.c.\right\},
\label{HIwidegap}
\end{eqnarray}
where the intraband contributions have been rewritten with the aid of the intraband current matrix element 
${\bf j}_{s\tau\bk}=-\frac{e}{\hbar}\nabla_{\bk}\epsilon_{s\tau\bk}$. 
From this simplified LM Hamiltonian, Eq. (\ref{HIwidegap}), one recognizes clearly that the valence-to-conduction-band 
excitations at the $K^\pm$ valley require the absorption of a $\sigma^\mp$ polarized photon, as has been 
shown by Xiao et al.\cite{Xiao2012}.

The optically induced current can be obtained
from the LM Hamiltonian via ${\bf j}=-c\langle \frac{\delta H_I}{\delta {\bf A}}\rangle$ and contains 
both intra- and interband transitions. From Eq. (\ref{HIwidegap}) it is clear that a finite macroscopic
interband current requires an even part of the microscopic 
transition amplitudes $P_{s\tau\bk}={\rm e}^{-i\tau\theta_{\bk}}\langle\nu^\dagger_{s\tau\bk}c_{s\tau\bk}\rangle$,
while 
intraband transitions contribute to the macroscopic current only if  
 $g_{s\tau\bk}=\langle\nu^\dagger_{s\tau\bk}\nu_{s\tau\bk}\rangle-\langle c^\dagger_{s\tau\bk}c_{s\tau\bk}\rangle$
contains an odd part.

The microscopic 
transition amplitudes $P_{s\tau\bk}$ 
and the Pauli blocking factor $g_{s\tau\bk}$ can be computed from the Heisenberg equations of motion yielding the semiconductor Bloch 
equations (SBE)\cite{HaugKoch}. Using Eq. (\ref{HIwidegap}) and the standard Coulomb-interaction Hamiltonian, one obtains
\begin{eqnarray}
\label{SBEP}
i\hbar \partial_t P_{s\tau\bk}&=&2\left(\Sigma_{s\tau\bk}-\frac{1}{c}{\bf A}\cdot{\bf j}_{s\tau\bk}\right)
P_{s\tau\bk}
\nonumber\\
&-&g_{s\tau\bk}\left(\sum_{\bk'}V_{\bk,\bk'}P_{s\tau\bk'}
-e\frac{v_F}{c}
A^{-\tau}\right),\\
i\hbar \partial_t g_{s\tau\bk}
&=&
2P^*_{s\tau\bk}\left(\sum_{\bk'}V_{\bk,\bk'}P_{s\tau\bk'}
-e\frac{v_F}{c}
A^{-\tau}\right)
-c.c.
\label{SBEf}
\end{eqnarray}
where  $\Sigma_{s\tau\bk}
=\epsilon_{s\tau\bk}+\frac{1}{2}\sum_{\bk'}V_{\bk,\bk'}g_{s\tau\bk'}$
denotes the renormalized single particle energy that  
includes the Coulomb renormalization of the band-gap.

\subsection{Regime of weak Coulomb coupling}

Assuming the noninteracting ground state as the initial state before optical excitation,
the linear response SBE for the polarization is given by
\begin{eqnarray}
\label{SBENI}
i\hbar \partial_t P_{s\tau\bk}&=&2\Sigma_{s\tau\bk} P_{s\tau\bk}-\sum_{\bk'}V_{\bk,\bk'}P_{s\tau\bk'}
-e\frac{v_F}{c} A^{-\tau}.
\end{eqnarray}
Since the populations are at least second order in the exciting field, we use $g_{s\tau\bk}=1$.
The expansion into the eigenstates of the Wannier equation with the relativistic 
single-particle dispersion
\begin{eqnarray}
2\Sigma_{s\tau\bk}\phi_{\mu}^{s\tau}(\bk)-\sum_{\bk'}V_{\bk,\bk'}
\phi_{\mu}^{s\tau}(\bk')
=
E_{\mu}^{s\tau}\phi_{\mu}^{s\tau}(\bk)
\label{wannier}
\end{eqnarray}
yields
\begin{eqnarray}
\chi^\sigma(\omega)&=&-\frac{e^2 v_F^2}{\omega^2}\sum_{s,\mu}\frac{  {\cal{F}}^{s,-\sigma}_\mu}
{\hbar(\omega+i\gamma)-E_{\mu}^{s,-\sigma}}\nonumber\\
&+&\frac{e^2 v_F^2}{\omega^2}\sum_{s,\mu}\frac{  {\cal{F}}^{s,\sigma}_\mu}
{\hbar(\omega+i\gamma)+E_{\mu}^{s,\sigma}} .
\label{Elliot}
\end{eqnarray}
for the linear susceptibility $\chi$, describing the optical response via
 \begin{eqnarray}
j^\sigma=\frac{\omega^2}{c}\chi^\sigma(\omega)A^ \sigma.
\end{eqnarray}
Here, the oscillator strength is given as
${\cal{F}}^{s,\tau}_\mu=\left|\sum_{\bk}\phi_{\mu}^{s,\tau}(\bk)\right|^2\label{oscillatorstrength}$.

The fully relativistic Wannier equation (\ref{wannier}) corresponds to the $k$-space representation
of the Dirac Coulomb problem. 
In real space, the wave 
functions are two-component spinors\cite{Pereira2008,Rodin2013}
\[
\P^{s\tau}_{jn}(\br)=\left(
\begin{array}{c}{\rm e}^{i( j-\tau/2)\varphi}A^{s\tau}_{j n}(r)\\ {\rm e}^{i( j+\tau/2)\varphi}B^{s\tau}_{j n}(r)\end{array}
\right)
\]
obeying the wave equation
\begin{equation}
 \left(2 v_F\hat{\bf \sigma}\cdot\bp+ \Delta_{s\tau}
\hat\sigma_z-\frac{e^2}{r}\right)\Psi_{jn}^{s\tau}(\br)=E_{jn}^{s\tau}\Psi_{jn}^{s\tau}(\br).
\label{wannierrel}
\end{equation}
Here $j$ is the eigenvalue of the total angular momentum $\hat L_z+\frac{\tau}{2}\hat\sigma_z$,
and  $n$ is the principle quantum number. 
We can identify the set of quantum numbers $\{\mu\}=(j,n)$ such that, e.g., an $s$
-type wave function has $j=\pm 1/2$.
The eigenfunctions $\phi_{jn}^{s\tau}$ of Eq. (\ref{wannier}) correspond to the solutions of Eq. (\ref{wannierrel}) with positive energy and are given by
\begin{eqnarray}
\phi_{jn}^{s\tau}(\bk)&=&\sqrt{\frac{\epsilon_{s\tau\bk}+\frac{\Delta_{s\tau}}{2}}{2\epsilon_{s\tau\bk}}}{\rm e}^{i( j-\tau/2)\theta_{\bk}}A^{s\tau}_{j n}(k)\nonumber\\
&+&\sqrt{\frac{\epsilon_{s\tau\bk}-\frac{\Delta_{s\tau}}{2}}{2\epsilon_{s\tau\bk}}}{\rm e}^{i( j+\tau/2)\theta_{\bk}}B^{s\tau}_{j n}(k)\nonumber\\
&\approx&
{\rm e}^{i( j-\tau/2)\theta_{\bk}}A^{s\tau}_{j n}(k),
\end{eqnarray}
while the solutions corresponding to negative energies are of no physical relevance here.

For the 2D Coulomb interaction potential, the eigenvalues of the 
relativistic hydrogen problem are given by\cite{Pereira2008,Rodin2013}
\[
E^{s\tau}_{nj}= \Delta_{s\tau}
\left\{
\frac{n+\sqrt{j^2-\left(\frac{\alpha}{2}\right)^2}}
{\sqrt{\left(\frac{\alpha}{2}\right)^2+\left(n+\sqrt{j^2-\left(\frac{\alpha}{2}\right)^2}\right)^2}}\right\}
\]
with $n=0,1,\dots$ and
 $j=\pm1/2,\pm3/2,..$. Here, $\alpha=e^2/\kappa\hbar v_F$ is the coupling constant and $\kappa$ is the effective dielectric constant of the environment. The eigenvalues are real for those $j$-values where $j^2-\left(\frac{\alpha}{2}\right)^2\ge 0$. For small values of $\alpha$, the eigenvalue spectrum reduces to that of the nonrelativistic Rydberg series 
\[
E_{nj}^{s\tau} = \Delta_{s\tau}-\alpha^2\Delta_{s\tau}/8(n+|j|)^2
= \Delta_{s\tau}-\alpha^2\Delta_{s\tau}/8(N-1/2)^2 
\]
with $N=n+|j|+1/2=1,2,\dots$.

From the expression (\ref{Elliot}) for the linear susceptibility, one can derive the exciton spectrum and 
optical selection rules for dipole allowed optical transitions. 
At the poles of Eq. (\ref{Elliot}), the reflection and absorption spectra display resonances.
The oscillator strength is a measure for the transition probability from the noninteracting groundstate to the
excitonic state with quantum number $\mu$.

Thus, we see that only excitonic transitions with $s$-type orbital angular momentum are dipole allowed.
Furthermore, the resonant contributions 
for a given circular polarization component $\sigma$ stem from the $K^{-\sigma}$ valley, while 
the nonresonant contributions stem from the $K^{\sigma}$ valley. This opens the possibility of a 
valley selective excitation with circularly polarized light. Moreover,
the dependence of the resonance energies $E_\mu^{s,\sigma}$ on the product of the spin and 
valley index couples the valley dynamics to the spin dynamics. Both effects have been predicted in \onlinecite{Xiao2012,Molina-Sanchez2013,
Cao2012,Zeng2012,Mak2012,Zeng2013} and demonstrated experimentally for various monolayers of TMDs.

\subsection{Regime of Strong Coulomb Coupling}

Interestingly, in the regime of strong Coulomb coupling characterized by $\alpha>1$, the optical spectrum contains no $s$-type wave functions with $j=\pm 1/2$. In this case, total angular momentum conservation cannot be fulfilled for optical transitions from the noninteracting groundstate such that no one-photon transitions should be possible.

To check if any of the TMD systems can be in the regime of strong Coulomb coupling, we 
use the material parameters given in Ref. \onlinecite{Xiao2012}. Thus, we obtain the nominal values
$\alpha= 3.29  /\kappa$ for WS$_2$, $\alpha= 3.66  /\kappa$ for WSe$_2$ and $\alpha=  4.11 /\kappa$ for MoS$_2$ 
respectively.
In free standing monolayers of TMDs, these values are very large and even on top of a 
typical substrate like SiO$_2$ with $\kappa=(3.9+1)/2$,  the effective coupling exceeds the critical value.
Thus, the assumption of a noninteracting ground state predicts {\it the absence of the one-photon optical transitions} for all of these systems, clearly contradicting the experimental findings.

However, in the regime of strong Coulomb coupling, the real part of the lowest exciton energy $E_{0\pm\frac{1}{2}}$ vanishes, i.e., it is degenerate with the noninteracting groundstate and the system transitions into
an excitonic insulator state. This excitonic insulator state is a BEC-like condensate of excitons, i.e.
a coherent superposition of the noninteracting groundstate and all exciton states with $\Re\left[E_\mu\right]= 0$\cite{Keldysh1965,Keldysh1968,Jerome1967,Halperin1968,Littlewood1996}.
This state exhibits a
static interband polarization $P_{s\tau\bk}^{\left[0\right]}=\Omega_{s\tau\bk}^{\left[0\right]}/2{\cal E}_{s\tau\bk}$  
and static blocking factor $g_{s\tau\bk}^{\left[0\right]}=\Sigma_{s\tau\bk}^{\left[0\right]}/{\cal E}_{s\tau\bk}$  
that can be obtained as nontrivial solutions of the gap equations\cite{stroucken2011,stroucken2013}
\begin{eqnarray}
 \Omega_{s\tau\bk}^{\left[0\right]}&=&\frac{1}{2}\sum_{\bk'}V_{\bk,\bk'}\frac{\Omega_{s\tau\bk'}^{\left[0\right]}}{{\cal E}_{s\tau\bk'}},\\
 \Sigma_{s\tau\bk}^{\left[0\right]}&=&\epsilon_{s\tau\bk}+\frac{1}{2}\sum_{\bk'}V_{\bk,\bk'}
 \frac{\Sigma_{s\tau\bk'}^{\left[0\right]}}{{\cal E}_{s\tau\bk'}} \, .
 \end{eqnarray}
Here, ${\cal E}_{s\tau\bk}=\sqrt{ \left.\Omega_{s\tau\bk}^{\left[0\right]}\right.^2+
\left.\Sigma_{s\tau\bk}^{\left[0\right]}\right.^2}$ is the 
dispersion of the Bogoliubov bands.

The static groundstate polarizations and populations induce a coupling between the linearized equations for the
optical polarization and populations and add an additional source term to the equation of motion for the optical
interband transition amplitude. Explicitely, we have

\begin{eqnarray}
i\hbar\partial_t P_{s\tau\bk}^{\left[1\right]}&=&2\Sigma_{s\tau\bk}^{\left[0\right]} P^{\left[1\right]}_{s\tau\bk}-
g_{s\tau\bk}^{\left[0\right]}\sum_{\bk'}V_{\bk,\bk'}P^{\left[1\right]}_{s\tau\bk'}\nonumber\\
&+&2 P_{s\tau\bk}^{\left[0\right]}\sum_{\bk'}V_{\bk,\bk'}g^{\left[1\right]}_{s\tau\bk'}
-
\Omega^{\left[0\right]}_{s\tau\bk'}g_{s\tau\bk}^{\left[1\right]}
\nonumber\\
&-&e\frac{v_F}{c}g_{s\tau\bk}^{\left[0\right]} A^{-\tau} 
-\frac{2}{c}{\bf A}\cdot{\bf j}_{s\tau\bk} P_{s\tau\bk}^{\left[0\right]},
\label{SBEPEx}
\\
i\hbar \partial_t g_{s\tau\bk}^{\left[1\right]}
&=&
2 \Omega^{\left[0\right]}_{s\tau\bk'} P_{s\tau\bk}^{\left[1\right]*}
+2 P_{s\tau\bk}^{\left[0\right]*}\sum_{\bk'}V_{\bk,\bk'}P^{\left[1\right]}_{s\tau\bk'}-c.c.
\nonumber\\
\label{SBEfEx}
\end{eqnarray}

As the additional source term is proportional to the intraband current matrixelement with odd parity, 
it mixes states with different 
parity. Though the odd terms of the linear interband polarization do not couple to the optical field directly,
they serve as source terms for the optically induced intraband current.
Eqs. (\ref{SBEPEx}) and (\ref{SBEfEx}) can be decoupled by the transformation
\begin{eqnarray*}
\label{trafo}
 \Pi_{s\tau\bk}^{\left[1\right]}&=&\frac{{\cal E}_{s\tau\bk}+\Sigma_{s\tau\bk}^{\left[0\right]}}{{\cal E}_{s\tau\bk}}P_{s\tau\bk}^{\left[1\right]}
-\frac{{\cal E}_{s\tau\bk}-\Sigma_{s\tau\bk}^{\left[0\right]}}{{\cal E}_{s\tau\bk}}P_{s\tau\bk}^{\left[1\right]*}
-\frac{\Omega_{s\tau\bk}^{\left[0\right]}}{{\cal E}_{s\tau\bk}}g_{s\tau\bk}^{\left[1\right]},\\
\Gamma_{s\tau\bk}^{\left[1\right]}&=&\frac{\Omega_{s\tau\bk}^{\left[0\right]}}{{\cal E}_{s\tau\bk}}\left(
P_{s\tau\bk}^{\left[1\right]}+P_{s\tau\bk}^{\left[1\right]*}\right)+
\frac{\Sigma_{s\tau\bk}^{\left[0\right]}}{{\cal E}_{s\tau\bk}}
g_{s\tau\bk}^{\left[1\right]}, 
 \end{eqnarray*}
yielding
\begin{eqnarray}
i\hbar\partial_t \Pi_{s\tau\bk}^{\left[1\right]}&=&2{\cal E}_{s\tau\bk}^{\left[0\right]} 
-\sum_{\bk'}V_{\bk,\bk'}\Pi^{\left[1\right]}_{s\tau\bk'}\nonumber\\
&-&2e\frac{v_F}{c}g_{s\tau\bk}^{\left[0\right]} A^{-\tau} 
-\frac{4}{c}{\bf A}\cdot{\bf j}_{s\tau\bk} P_{s\tau\bk}^{\left[0\right]},
\label{SBEBog}
\\
i\hbar \partial_t \Gamma_{s\tau\bk}^{\left[1\right]}
&=&0.
\end{eqnarray}
Hence, the solutions of Eq. (\ref{SBEBog}) can be expanded in terms of the eigenfunctions of the Bogoliubov-Wannier equation
\begin{equation}
2{\cal E}_{s\tau\bk}\psi_{\mu}^{s\tau}(\bk)-\sum_{\bk'}V_{\bk,\bk'}
\psi_{\mu}^{s\tau}(\bk')=E^{s\tau}_{\mu}\psi_{\mu}^{s\tau}(\bk'),
\label{generalizedwannier}
\end{equation}
where the dispersion of the noninteracting bands has been replaced by the Bogoliubov quasi-particle dispersion.
The groundstate polarization obeys
$2{\cal E}_{s\tau\bk}P_{s\tau\bk}^{\left[0\right]}-\sum_{\bk'}V_{\bk,\bk'}
P_{s\tau\bk'}^{\left[0\right]}=0$ such that it
can be expanded into the eigenfunctions of the Bogoliubov-Wannier equation with the eigenvalue $E^{s\tau}_{\mu}=0$.
Hence, we make the ansatz 
\begin{eqnarray}
 P_{s\tau\bk}^{\left[0\right]}&=&\sum_{\mu,E_\mu=0}
 P_{s\tau\mu}^{\left[0\right]} \psi^{s\tau}_\mu(\bk),\\
 \Pi_{s\tau\bk}^{\left[1\right]}&=&\sum_{\mu}\Pi_{s\tau\mu}^{\left[1\right]}\psi^{s\tau}_\mu(\bk),
\end{eqnarray}
Inserting the ansatz into Eq. (\ref{SBEBog}) yields
\begin{equation}
  \Pi_{s\tau\mu}^{\left[1\right]}(\omega)=
 -2e\frac{v_F}{c}A^{-\tau}\frac{(\gamma_\mu^{s\tau})^*}{\hbar\omega-E_\mu}
 -\frac{4}{c}\frac{{\bf A}\cdot{\bf j}^{s\tau}_\mu}{\hbar\omega-E_\mu}.
 \label{Pilambda}
\end{equation}
Here, $\gamma_\mu^{s\tau}=\sum_{\bk}g_{s\tau\bk}^{\left[0\right]}\psi_\mu(\bk)$
is the coupling strength of the
macroscopic interband polarization to the optical field. 
For an $s$-type groundstate ($1<\alpha< 3$), the
populations also have spherical symmetry and the coupling strength vanishes for all
non $s$-type excition states, i.e. for all real exciton  resonances.
Generally, if $\alpha<(2n+1)/2$, the groundstate 
populations are composed of exciton states with $|j|\le(2n+1)/2$, while real exciton resonances require
$|j|\ge(2n+1)/2$. Thus, in the strong Coulomb coupling regime, the first term on the RHS of Eq. (\ref{Pilambda})
 vanishes.

Reversing the transformation \ref{trafo}, we can compute the optical susceptibility for 
the excitonic insulator state
\begin{eqnarray}
\chi^\sigma(\omega)_{EI}&=&-\frac{1}{\omega^2}\sum_{s,\mu}
\frac{  
|{\bf j}^{s-\sigma}_\mu|^2
}
{\hbar(\omega+i\gamma)-E_{\mu}^{s,-\sigma}}
\nonumber\\
&+&\frac{1}{\omega^2}\sum_{s,\mu}
-\frac{ |{\bf j}^{s\sigma}_\mu|^2}
{\hbar(\omega+i\gamma)+E_{\mu}^{s,\sigma}}
\label{ElliotEI}.
\end{eqnarray}
Here, all contributions have intraband character.


As can be recognized from Eq. (\ref{ElliotEI}), the poles of the linear susceptibility for the excitonic insulator state occur at the spectrum of the exciton Hamiltonian with Bogoliubov dispersion. The dominant effect of the exciton condensation on the single quasi-particle dispersion is a renormalization of the gap. Hence, the spectrum can be obtained in very good approximation from  Eq. \ref{wannierrel}, provided one uses the appropriate renormalized gap.
The major difference as compared to the optical susceptibility, Eq.(\ref{Elliot}),
for the uncorrelated groundstate is the oscillator strength, which is now given by the
intraband current matrix element 
between the excited exciton state and the excitonic groundstate
\[
{\bf j}^{s\tau}_\mu=\sum_{\bk}\left.\psi^{s\tau}_{\mu}\right.^*(\bk){\bf j}_{s\tau\bk} P_{s\tau\bk}^{\left[0\right]}\equiv \langle \mu s\tau|
{\bf j}|0\rangle.
\]
This matrix element connects that part of the groundstate with
largest $|j|\le\alpha/2$ to the excited states with $|j'|=|j|+1>\alpha/2$
such that dipole allowed transitions
increase (decrease) the total angular momentum by one. Explicitly,
 in the regime with $1<\alpha< 3$, optical transitions 
connect the $s$-type excitonic groundstate $j=\tau /2, l=0$ with the excited excitonic states $j'=3\tau/2,l=\tau$, 
corresponding to $p$-like states.
Thus, combining the results of the weakly and strongly interacting regimes,  the modified optical selection rules can be
summarized by  the condition 
\begin{equation}
(2(|l|-1)< \alpha < (2|l|+1)
\end{equation}
for the optically active states.

\section{Analysis of Experimental Observations}

To test our general theory, we compare its predictions with the observations of several recent experiments on 
WS$_2$ \cite{Chernikov2014,Zhu2014,Ye2014} and WSe$_2$\cite{Wang2014}. 
As commonly adopted, 
we refer to A and B excitons for the lower and higher direct exciton transitions with a spin and valley 
index combination $s\tau=1,-1$ respectively.

While the effective fine structure constant can be extracted from the 
tight binding parameters and the dielectric screening,
to compute the energetic positions of the bright states requires 
the additional knowledge of the Coulombic bandgap renormalization.
Whereas the  bandgap renormalization diverges for an ideal, strictly 2D Coulomb potential, 
it is finite for any real system where the Coulomb matrix elements must be 
evaluated  from the gap equations with the atomic orbitals used to represent the 
pseudo spinor space in Eq. (\ref{Hamiltonian}). This leads to a quasi-2D Coulomb interaction
\begin{equation}
V(\bq)=\frac{2\pi e^2}{\cal{A}\kappa}\frac{F(qd)}{q},
\end{equation}
where $\cal{A}$ is the normalization area
and $F(qd)$ is a 
monotonically decreasing form factor with $F(0)=1$ 
and $\lim_{x\rightarrow\infty}xF(x)=0$.
The form factor leads  to a regularization of the $1/r$ 
singularity at small distances and the parameter $d$ can be interpreted as the effective thickness of the 
monolayer.  Since its explicit expression depends on the specific  atomic orbitals contributing to the 
valence and conduction band,  we do not calculate the bandgap 
but rather  use it as a fit parameter.  

\subsection{Linear Reflection experiments}

\begin{figure}
\centerline{\includegraphics[width=7.8cm]{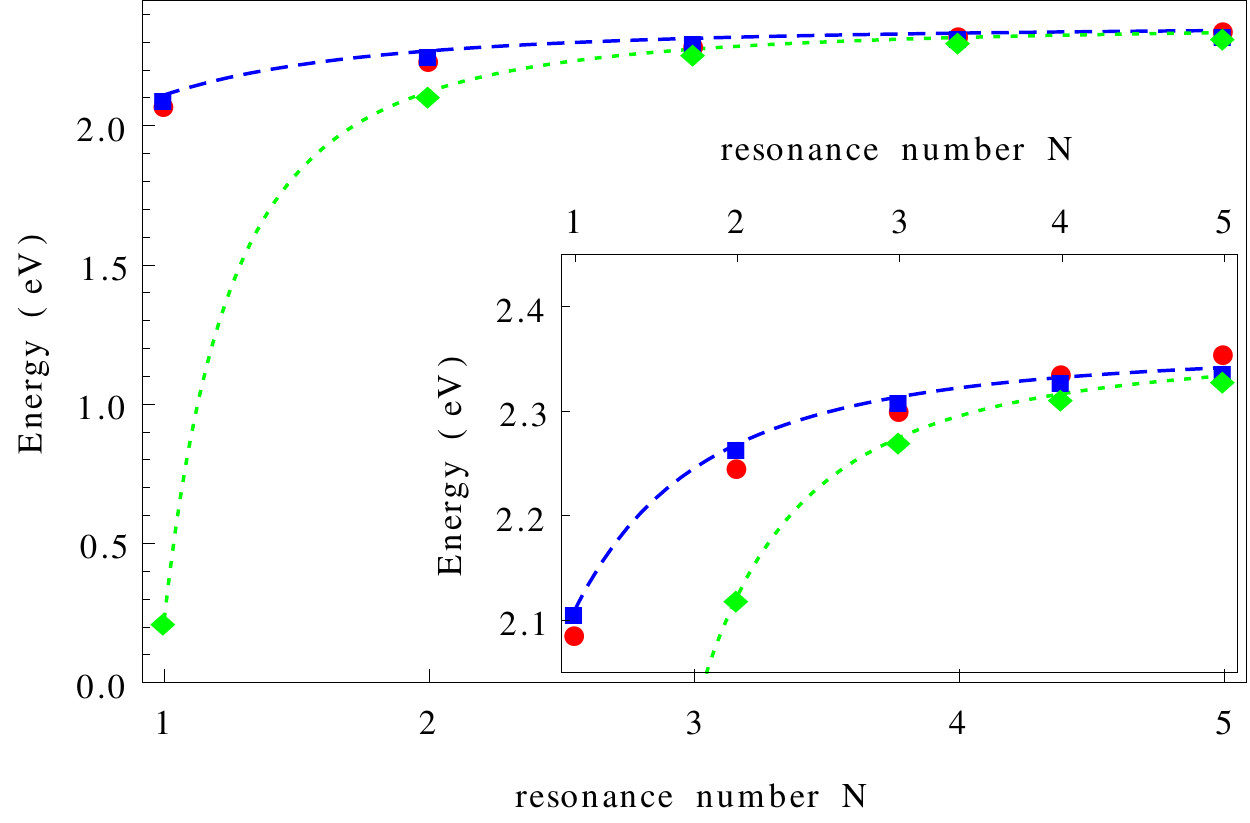}}
\caption{(color online) Spectral position of the energetically lowest bright excitonic transitions with $j=\pm 3/2$ in WS$_2$. 
The blue squares are the computed results for p-type excitons, the
red dots show the experimental data taken from \cite{Chernikov2014}, and the green diamonds indicate the theoretical 
predictions if one assumes 
 $s$- type  bright states, respectively. The dashed and dotted lines are guides to the eye. 
 The x-axis label denotes the number $M$ of the experimentally observed bright exciton resonances and is related to the priniple quantum number  
 via $N=M-1$.  
The inset to the figure shows the results for the energetically higher states with a finer energy resolution.}
\label{Comparisontoexperiment}
\end{figure}

We  start with the analysis of the bright
excitonic states in WS$_2$ on SiO$_2$ corresponding to 
the lowest A exciton with spin and valley combination $s\tau=1$\cite{Chernikov2014}. 
With the material parameters for WS$_2$  $t=1.37$eV,  $2\lambda=0.43$ eV, $a=3.197$\r{A} 
given in Ref. \cite{Xiao2012},
an effective dielectric constant $\kappa=(1+3.9)/2$ for the SiO$_2$ substrate, and using $E_{gap}=2.36$ eV,
we find a binding energy
 of $0.25$ eV for the lowest bright exciton transition with $(n,j)=(0,\pm3/2)$.
Using $E_{gap}=2.36$ eV, we present in Fig. \ref{Comparisontoexperiment}
the computed result for the spectral position of the lowest five bright exciton transitions. 
The red diamonds show the experimental data points taken from Ref. \onlinecite{Chernikov2014}. For comparison,
we also show the predicted values assuming nonrelativistic bright $s$-states. (Note that the $j=\pm1/2$ 
series gives complex eigenvalues of the relativistic wave equation and thus cannot be used for comparison.)
Our computed results show remarkably good agreement with the experimental observations given the fact that only 
the effective bandgap was used as a single fit parameter.
In contrast, the results based on the assumption that the series has 
$s$-type character  are completely off the scale, in particular for the energetically 
lowest states.

Assuming that the observed spin splitting corresponds to the difference in the renormalized
bandgaps, we obtain $E^B_{gap}=2.79eV$ which can be used to compute the spectral 
positions  of the B-excitons. 
The ratio of the binding energy is fixed by the ratio of the renormalized gaps,
i.e. $E^B_0/E^A_0=E^B_{gap}/E^A_{gap}$, giving 
a binding energy of $0.30$ eV for the lowest bright B exciton. Thus the $2p$ resonance is 
predicted to occur at $E^B_{2p}=2.49$ eV, which is also in very good agreement with the experimental
findings\cite{Chernikov2014}. 
In table \ref{Table1}, a list of the lowest theoretically predicted resonance positions of both A and B exciton is given.

\begingroup
\squeezetable
\begin{table}
\begin{tabular}{| c |c c c c c c c c  |c| c |}
\hline
 &  $E^A_2$ &  $E^A_{3}$ & $E^A_{4}$ & $ E^A_{G}$ & $ E^B_{2}$ & $  E^B_{3}$  &$E^B_4$& $ E^B_{G}$& method& Ref.\\
\hline
\hline
\multirow{6}{*}{WS$_2 $}
                         &2.11& 2.27 &2.31 & 2.36& 2.49 & 2.68 & 2.74&  2.79&theory&\\
                         &2.09 & 2.25 & 2.3  &          & 2.5&    &   &     &linear reflection&\cite{Chernikov2014}\\
                         &2.02&         &         &          &2.4&     &     &   &linear absorption&{\cite{Zhu2014}}\\
                         &       &         &         &          &2.4&2.6&  &2.73&TP-PLE& \cite{Zhu2014}\\
                         &2.04&         &         &          &2.45&    &    &   & linear reflection&   \cite{Ye2014}\\
                         &x     & 2.25 & 2.29 & 2.35 &2.46&    &    &  & TP-PLE&\cite{Ye2014}\\
\hline
\multirow{4}{*}{WSe$_2 $}
                          & 1.75 & 1.91 &1.96 & 2.01&2.15  &2.35 &2.41 &2.47&theory&\\
                          &1.75&      & & & & &&  &PLE&\cite{Wang2014}\\
                          &x     &1.9 & & &x & &&&TP-PLE&\cite{Wang2014}\\
                          &1.75&1.91&1.98&2.02&2.17&2.3&& &SHG&\cite{Wang2014}\\
\hline
\end{tabular}
\caption{ Excitonic resonance positions extracted from experimental data together with the  
theoretical predictions where the gap of the A-exciton has been used as single fit parameter to match the 
experimental data. For WS$_2$, the band gap has been fitted to match the data of Ref. \onlinecite{Chernikov2014} (see Fig. 1 and the discussion in the text). For WSe$_2$, the lowest bright resonance position has been fitted to the data of Ref. \onlinecite{Wang2014}.  The data of Ref. \onlinecite{Zhu2014} correspond to room temperature and show a red shift of approximately $0.1$ eV as compared to the linear reflection spectra at 10K. \label{Table1}. The labeling corresponds to the main quantum number $ N$ of the $ Nth$ excitonic level, which is usually used in semiconductor optics.
 }
\end{table}
\endgroup
\subsection{Two Photon Experiments}

\begin{figure}
\centerline{\includegraphics[width=7.6cm]{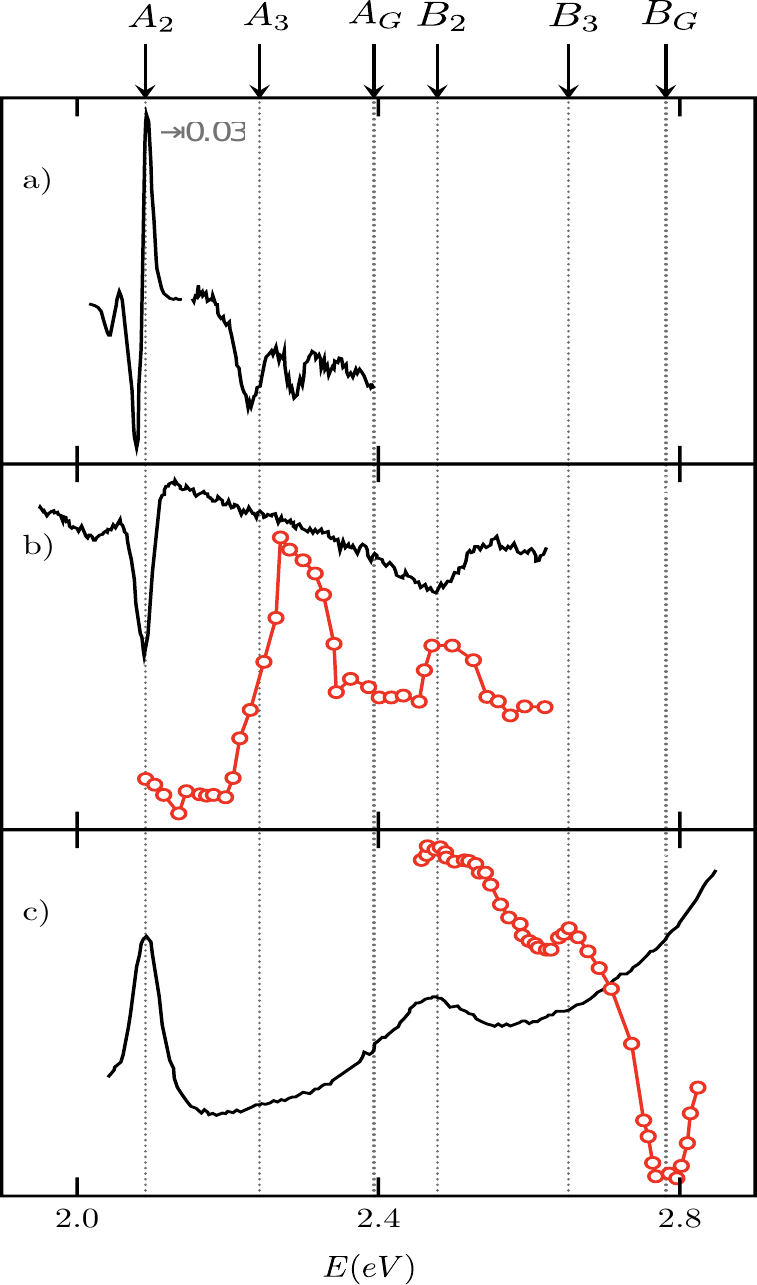}}
\caption{(color online) 
Comparsion of one and two photon spectra of WS$_2$ measured by different groups. Part a) shows the 
deravative of the reflection contrast  
measured by Chernikov et al.\cite{Chernikov2014}, part b) the the linear reflection contrast (black line) 
together with the TP-PLE spectrum (red dots)
measured by Ye at al. \cite{Ye2014}, and part c) the linear absorption (black line) and TP-PLE (red dots) measured by Zhu et al.
\cite{Zhu2014}. To compensate for temperature dependent shifts of the band gaps, we shifted the data 
such that the lowest resonance of 
the A excitons are alligned. On the upper axis, we assigned the quantum numbers assuming bright $p$ states. 
The onsets of the band gaps are denoted by $A_G$ and $B_G$. The dotted 
line are guides to the eye and show the energetic position of some selected states.}
\label{spectraWS2experimental}
\end{figure}

To provide further evidence for the theoretically predicted selection rules, 
we also analyze recent experiments probing dark states
by two-photon spectroscopy which can only access excitonic states with a parity opposite to
those excited by a single photon process\cite{Zhu2014,Ye2014,Wang2014}. In Ref. \onlinecite{Zhu2014,Ye2014}, linear reflection and TP-PLE have been performed on WS$_2$ flakes 
having nominally the same material parameters as those in  Ref.\onlinecite{Chernikov2014}.  
The resulting experimental data are shown in Fig. \ref{spectraWS2experimental}.

In Ref. \onlinecite{Zhu2014}, the linear reflection has been measured for different temperatures, 
showing two clearly recognizable peaks for $T=10$K at $2.1$ eV
and $2.5$ eV, respectively. These peaks agree well with the dominant peaks in the reflection 
spectrum of Ref. \onlinecite{Chernikov2014} and can be assigned to the  A and B exciton respectively. 
At room temperature, both peaks experience a red shift of approximately $0.1$ eV,  most likely 
resulting from a temperature dependent bandgap renormalization. The TP-PLE measurements have been
performed at room temperature and cover the spectral range
between $2.4$ and $3.0$ eV. Within this spectral range two peaks in the TP-PLE spectrum are observed
at approximately $2.4$ and $2.6$ eV, and a significant gap at approximately $2.73$ eV.
In Ref.  \onlinecite{Ye2014}, the A and B excitonic resonances occur at $ 2.04$ and $2.45$eV in the 
linear reflection spectrum respectively. Thus, the energetic separation of the two dominant peaks is 
in good agreement with Refs. \onlinecite{Chernikov2014} and  \onlinecite{Zhu2014}. The small deviations in the 
absolute values are most likely due to different experimental conditions. The TP-PLE spectrum covers 
the range between $1.9$ and $2.6$ eV and thus complements the measurements of Ref. \onlinecite{Zhu2014}.
Within this energy range, the TP-PLE spectrum exhibits pronounced maxima between $2.25-2.29$ eV and at $2.45$ eV.
Remarkably, both experiments observe a large TP-PLE signal resonant with the lowest bright transition 
of the B exciton, whereas
the TP-PLE in Ref. \onlinecite{Ye2014} shows no significant signal at the lowest bright A exciton.

In Ref. \onlinecite{Wang2014},
TP-PLE measurements have been performed in conjunction 
with SHG experiments on WSe$_2$. This material system has very similar properties as WS$_2$, 
in particular it has the same symmetries and similar spin-splitting of the valence bands. Using PLE, 
the resonance position of the A exciton is determined as $1.75$ eV, whereas in TP-PLE,  
the dominant peak occurs at approximately 1.9 eV.  In agreement with the observation in
Ref. \onlinecite{Ye2014}, the TP-PLE spectrum shows no significant signal at the transition 
energy of the lowest bright  A exciton. In contradiction to the results of Ref. \onlinecite{Ye2014},
the TP-PLE spectrum does not show a significant signal at the transition 
energy of the lowest bright B exciton.
In contrast, the SHG spectrum exhibits two dominant peaks at
$1.75$ eV, i.e, at the lowest bright resonance,  and at $2.16$ eV, and less prominent peaks 
at $1.9$, $2.02$ and $2.3$ eV. 

In Refs. \onlinecite{Zhu2014,Ye2014,Wang2014}, the experiments have been analyzed under 
the assumption of bright $s$-excitons. This assumption is seemingly supported by the absence of a TP-PLE signal 
at the resonance frequency of the lowest bright A exciton transition\cite{Ye2014}, 
and at the lowest A and B exciton\cite{Wang2014} respectively. Thus,  the resonances
observed in TP-PLE spectra are assigned to $p$-like states. Since the lowest $p$-state should be energetically 
well above the lowest $s$- state, all resonances observed in the TP-PLE spectra of WS$_2$  
are therefore assigned to excited states of the A exciton, 
including those resonant with the lowest bright B exciton and above.
This assignement directly contradicts the Rydberg series observed by 
Chernikov et al.\cite{Chernikov2014}, where all excited states and the 
onset of the band gap of the A-exciton are found {\it below} the B- exciton. 

Remarkably, all the above mentioned experimental observations can be explained quite naturally by the selection rules predicted by our theory. 
In the strongly interacting regime, the lowest bright exciton state is $p$-type and nondegerate, as $s$-type wave 
functions do not exist. The higher states  are at least two-fold degenerate, and a two-photon absorption excites
a $d$-type excited state. Thus, if the groundstate is excitonic, the lowest excited state should  be dark in a 
two-photon experiment, exactly as is the case for the lowest $s$-exciton in a weakly interacting system.
However, if the coupling constant is slightly below the critical value, the $j=1/2$ states become bright and the
state with $(n,j)=(0,3/2)$ becomes (nearly) degenerate with the $(n,j)=(1,1/2)$ state, which is the first excited $s$-type state. Simultaneously, the $p$-state can only be reached via a two-photon transition, while the corresponding degenerate $s$-state would be bright. Thus, the observation of a TP-PLE signal at the resonance position of the lowest bright B-exciton in the WS$_2$
sample seems to indicate that the WS$_2$ sample is at the edge of an excitonic insulator, with the lowest B$1s$ exciton transition somewhere in the (far) infrared.

In Table   \ref{Table1}, we summarize the reported observations and compare our theoretically predicted resonance 
positions of WS$_2$ and WSe$_2$ 
with the experimentally available data. The theoretical values have been obtained using the material parameters of 
Ref. \cite{Xiao2012}, $\kappa=(1+3.9)/2$ with the band gap as a single fit parameter.
Note, that we assigned quantum numbers to the respective resonance positions that result from our 
theoretical assignment and do not necessarily correspond to the labeling in the respective original publications. 
Resonances that could not be resolved by the specific experiment remain as empty spaces in Table \ref{Table1} , while those resonances 
that could in principle occur in the experiment but are not observed are indicated by an 'x'. 
As can be recognized from the table, not only the agreement between theory and experiment is remarkable, 
but also the data obtained by the different experimental techniques 
on different samples is brought into almost perfect mutual agreement as soon as 
one adopts our interpretation in terms of optically active $p$-states.

\section{Discussion}

The results discussed in the previous sections are based on the model Hamiltonian (\ref{Hamiltonian}) treating the 
electrons and holes in the vicinity of the $K$-points as massive Dirac Fermions. The strength of this 
model Hamiltonian is its simplicity, allowing for an analytical solution of the exciton problem and thus providing 
insight into the optical properties of Coulomb-interacting chiral 
quasi-particles that are 
difficult to obtain numerically. The obvious disadvantage is that it contains several simplifications that
might restrict the application of the model to real material systems.
In the following, we give a brief discussion of the approximations underlying our analysis, 
and estimate its validity and restrictions.

Inherent to the massive Dirac model is a single-particle dispersion that has a perfect electron-hole symmetry. 
The effect of an eventual electron-hole asymmetry is twofold: It modifies the single-particle dispersion and therewith 
the homogeneous part of the Wannier equation and it alters the dipole matrix element, 
i.e. the light-matter interaction. The latter effect may weaken the 
optical selection rules, i.e. unlike for the case of a perfect electron-hole symmetry, one- and two-photon 
processes can address the same states, while the first effect yields a modified exciton spectrum.

For the exciton equation, the relevant quantity is the energy difference between the conduction and valence band.
In a conventional semiconductor with  parabolic bands, this leads to an exciton dispersion 
and binding energy that depend on the reduced mass of the electron-hole pair only. Adding a
term to the single-particle Hamiltonian of the form
\[
H_{as}^{s\tau}=\sum_{\bk} \P^\dagger_{a\tau\bk} k^2\left(
\begin{array}{cc}
\alpha_{s\tau}&0\\0&\beta_{s\tau}
\end{array}
\right)
\P_{s\tau\bk},
\]
that has been proposed in Refs. \onlinecite{Kormanyos2013,Rostami2013,Kormanyos2015} to account for different electron and 
hole masses within the chiral two-band model, it is straightforward to show
that, up to corrections of the order 
${\cal O}(k^4)$, the
exciton dispersion corresponds to that of a relativistic, symmetric electron-hole pair with a Fermi velocity 
$v_F^*=v_F\sqrt{\frac{m_0}{2m_r}}$ and effective mass $m^*=\Delta/2 v_F^{*2}$.
Here, $m_0=\Delta/2 v_F^2$ and $m_r$ is the reduced mass of the electron-hole pair. 
 It is worth to note that the so defined effective mass is not directly related to the electron and hole
masses but  describes the {\it pair} properties only.  
Using the parameters for WS$_2$ given in \cite{Kormanyos2015}, this gives $v_F^*=1.066 v_F$ 
and a nominal coupling constant $\alpha^*=3.08$ which is still well above the critical value. 
Thus, although a possible electron-hole asymmetry may lead to a violation of the optical selection rules,
the massive Fermion model is well suited to describe the exciton spectrum even for asymmetric electrons and holes, 
provided one uses the effective Fermi-velocity.

To obtain the spectral position of the excitons, we utilize 
the solutions of the resulting relativistic Wannier equation where we include only the substrate background 
screening but neglect all other screening effects. In general, screening effects 
arise both from filled valence bands that are not considered explicitely in the model Hamiltonian, and screening of the
conduction and valence band under consideration. The Coulomb matrix elements used to set up the interacting model 
Hamiltonian should contain only the first type of contributions which can be considered to lead to a constant  background 
dielectric constant. For a film of vanishing thickness $d$, this contribution can be neglected\cite{Wehling2011}.
The second type of contributions should then be computed selfconsistently, 
using the  Coulomb matrix elements screened by the background dielectric constant.

For the massive Dirac-Hamiltonian (\ref{Hamiltonian}), the RPA dielectric function of the TMD monolayer is given by
\cite{Rodin2013},
\[
 \epsilon(q)=\kappa\left(
 1+\frac{4}{\pi}\frac{1}{qa_B}P\left(\frac{\alpha}{2}qa_B\right)\right),
\]
where $\kappa$ is the effective dielectric screening of the substrate and $a_B=2\hbar v_F/\alpha\Delta $ is the $2D$ exciton Bohr radius. 
An analytical expression for $P(Q)$ is given  in Ref. \onlinecite{Rodin2013}.
In the nonrelativistic limit $\frac{\alpha}{2}qa_B\ll 1$, 
$P(Q)\approx \frac{\pi Q^2}{3}$,
yielding 
$\epsilon(q)=1+\frac{1}{3}\alpha^2 a_B q$. In this regime, the intrinsic screening corresponds to 
the anti-screening proposed by\cite{Keldysh1978,Cudazzo2011} with an anti-screening length 
$r_0= \frac{1}{3}\alpha^2 a_B$.
Using the nominal values of $\alpha$ in suspended TMDs, the anti screening length is
typically $\propto 1-1.5 \, a\approx 3-5$  \AA,  which corresponds roughly to 
the physical thickness of the monolayer. For supported samples,
the anti-screening length is reduced correspondingly.

The anti-screening is typical 
for a truly 2D-dielectric material and and weakens the strength of the Coulomb 
interaction on a length scale small compared to the anti-screening length.
Thus,  for the weakly bound excitons with a spatial extension large compared to the 2D-exciton Bohr radius, 
we expect the anti-screening effect to be of minor importance.
This expectation is indirectly confirmed by the excellent agreement of the experimentally 
observed resonance positions and our theoretical predictions.

At this point, we mention that the anti-screening model has also 
been employed to understand the observed non-hydrogenic series in TMDs 
based on an analysis in terms of optically bright $s$-states, using the anti-screening length as fit 
parameter\cite{Chernikov2014,Berkelbach2015}. For WS$_2$ on SiO$_2$,
reasonable agreement with the experimental observations could be obtained using $r_0=75$ \AA, which is much 
larger than the physical thickness.

In the  ultrarelativistic regime, 
$\frac{\alpha}{2}qa_B\gg 1$, $P(Q)\approx \pi^2Q/4$  and 
the  dielectric function $\epsilon(q)=1+2\alpha/\pi$ reduces to a constant, as in graphene. 
For large $q$-values, intrinsic screening therefore may play a significant role and eventually prevent the
phase transition into the excitonic state. 

Nevertheless,  the weakly bound states are not affected by this screening, nor by other deviations of the full
electron spectrum beyond the quadatic approximation. Thus,  the
excited states still have $|j|>\alpha/2$. Hence, assuming a noninteracting groundstate and a nominal value of 
$\alpha>1$, the excited exciton states should be dark and the only bright 
resonance in a linear optical experiment should correspond to the $1s$ exciton. 
Since the experiments do not observe this feature but rather show  
a whole series of excitonic resonances, this serves as a clear indication that the 
groundstate is indeed excitonic and the bright exciton resonances correspond to $p$-type wave functions with $|j|=3/2$.

\section{Conclusion}

In conclusion, we presented a general theory which assigns a $p$-like 
symmetry to the wave functions of the optically active excitonic transitions in recently 
investigated TMDs. This theory is supported by recently published experimental data,
including the observation of a series of resonances in the linear reflectrum spectrum of WS$_2$\cite{Chernikov2014}, 
two photon absorption in WS$_2$\cite{Zhu2014,Ye2014} and WSe$_2$\cite{Wang2014}.
This observation of optically active $p$-transitions is not only of crucial importance for the correct 
interpretation of experimental data, but impressively 
demonstrates that in the TMDs investigated here, the Coulomb 
interaction is strong enough to induce an excitonic ground state. 
The dipole allowed excited states of such an 
excitonic insulator correspond to intra-excitonic transitions and are governed 
by the corresponding optical selection rules. Even for systems in the weakly interacting 
regime without an excitonic ground state, the lowest excitonic transition should be in the THz or infrared regime 
and the observed optical resonances correspond to excited exciton levels with main quantum number $N \ge 2$.

Acknowledgements: This work is supported by the Sonderforschungsbereich 1083 funded by the Deutsche
Forschungsgemeinschaft. We thank the authors of Ref. \onlinecite{Chernikov2014} for 
sending us a preprint of their manuscript.




\end{document}